\documentclass[aps,pra,reprint,letterpaper,floatfix,superscriptaddress,amsmath,amssymb,nofootinbib]{revtex4-2}
\ifdefined\pdfminorversion\pdfminorversion=7\fi
\usepackage[utf8]{inputenc}
\usepackage{graphicx}
\usepackage{xcolor}
\usepackage[hidelinks]{hyperref}
\usepackage{cleveref}
\hypersetup{pdftitle={Intensity limits for phase-only holographic optical-tweezer arrays},pdfauthor={Anzhou Wen, Hengyu Lian, Kevin Yipu Wu, Maxwell F. Parsons, Praneeth Chakravarthula}}
\crefname{equation}{Eq.}{Eqs.}
\Crefname{equation}{Equation}{Equations}
\crefname{section}{Sec.}{Secs.}
\Crefname{section}{Section}{Sections}
\crefname{figure}{Fig.}{Figs.}
\Crefname{figure}{Figure}{Figures}
\renewcommand{\thesubsection}{\Alph{subsection}}

\makeatletter
\renewcommand{\p@subsection}{}
\renewcommand{\p@subsubsection}{\thesubsection\,}
\makeatother

\makeatletter
\def\toclevel@subsection{1}
\def\toclevel@subsubsection{2}
\makeatother
\begin{document}
\title{Intensity limits for phase-only holographic optical-tweezer arrays}
\author{Anzhou Wen}
\affiliation{Department of Computer Science, University of North Carolina at Chapel Hill, Chapel Hill, North Carolina 27599, USA}
\affiliation{Department of Applied Physical Sciences, University of North Carolina at Chapel Hill, Chapel Hill, North Carolina 27599, USA}
\author{Hengyu Lian}
\affiliation{Department of Computer Science, University of North Carolina at Chapel Hill, Chapel Hill, North Carolina 27599, USA}
\author{Kevin Yipu Wu}
\affiliation{\mbox{Department of Electrical and Computer Engineering, University of Washington, Seattle, Washington 98195, USA}}
\author{Maxwell F. Parsons}
\affiliation{\mbox{Department of Electrical and Computer Engineering, University of Washington, Seattle, Washington 98195, USA}}
\author{Praneeth Chakravarthula}
\email{cpk@cs.unc.edu}
\affiliation{Department of Computer Science, University of North Carolina at Chapel Hill, Chapel Hill, North Carolina 27599, USA}
\affiliation{Department of Applied Physical Sciences, University of North Carolina at Chapel Hill, Chapel Hill, North Carolina 27599, USA}
\begin{abstract}
Optical tweezers use tightly focused laser beams to trap and manipulate individual particles, enabling advances in quantum computing, drug discovery, and biological research. 
Scaling these technologies to large arrays requires delivering sufficient light to every trapping site while minimizing the total laser power required.
Here, we investigate the fundamental limits of how efficiently a phase-only hologram can distribute light across an array of optical traps.
We derive upper bounds on the minimum intensity across trapping sites, accounting for illumination, target geometry, and diffraction from spatial light modulator pixels. 
For arrays arranged in a single plane or across two axial planes, we show how these limits scale with the number of traps and SLM pixels, and establish conditions for focusing light in three dimensions.
Numerical experiments demonstrate that optimized holograms can closely approach these theoretical limits, suggesting that further improvements in phase optimization alone offer limited power savings.
Instead, our results reveal how the choice of illumination and trap geometry can improve light efficiency, providing practical guidelines for designing larger, more power-efficient optical-tweezer arrays.
\end{abstract}
\maketitle

Holographic optical tweezers divide a laser beam into programmable traps using a phase-only spatial light modulator (SLM) \cite{dufresne2001computer,kim2019large,dileonardo2007computer}. Each trap must receive enough light to hold a particle reliably. As arrays grow, the available laser power becomes a constraint. Because every trap must meet a minimum intensity requirement, whether for optical assembly \cite{shaw2017scanning} or scaling atom arrays \cite{holman2026trapping}, the weakest trap determines the incident power needed for the entire array. The central question is therefore how much intensity a phase-only hologram can deliver to its weakest trap.

Gerchberg--Saxton (GS) and weighted GS algorithms are widely used to design trapping holograms \cite{gerchberg1972practical,dileonardo2007computer,kim2019large}. Other methods use gradient-based optimization \cite{bowman2017high,nishimura2026optimization} or learned hologram generation \cite{ren2024creation}. Experimental feedback can further correct aberrations and improve focal shape, position, and intensity uniformity \cite{christen2025full,chew2024ultraprecise,tamura2016highly}. 
Although these methods improve the light delivered to the traps, an optimized hologram alone cannot tell us whether further phase optimization could substantially reduce the required laser power.

For a given illumination, target geometry and SLM, answering this question requires an upper bound on the weakest-trap intensity and a way to determine when it can be approached.
Prior work has established rigorous bounds on diffraction efficiency of phase-only holograms \cite{wyrowski1991upper,krackhardt1992upper}, developed optimal-grating theory for periodic beam division \cite{romero2007theory,romero2007theoryII}, and analyzed pixel diffraction and fill-factor losses in equal-intensity arrays \cite{arrizon1997efficiency,arrizon1999implementation}. 
We build on these foundations and address the question by deriving illumination- and geometry-dependent intensity bounds for pixelated phase-only holograms and establishing when these limits can be approached.

Specifically, we derive upper bounds on the weakest-trap intensity that account for illumination, target geometry, and SLM pixel diffraction, together with conditions under which this bound can be attained.
Our key step is to group aperture positions that contribute the same relative phase across the target lattice, allowing us to bound all target fields together.
We establish when the planar bound is attained in the large-array limit and when the same holograms form local three-dimensional intensity maxima near the targets. On a fixed SLM, comparisons with optimized holograms bound the remaining input-power savings from phase optimization.
Finally, we derive a matching limit for a family of two-plane arrays and show that distributing targets across depth can require less power than the specified family of planar layouts.

\begin{figure*}[t]
\centering
\includegraphics[width=\textwidth]{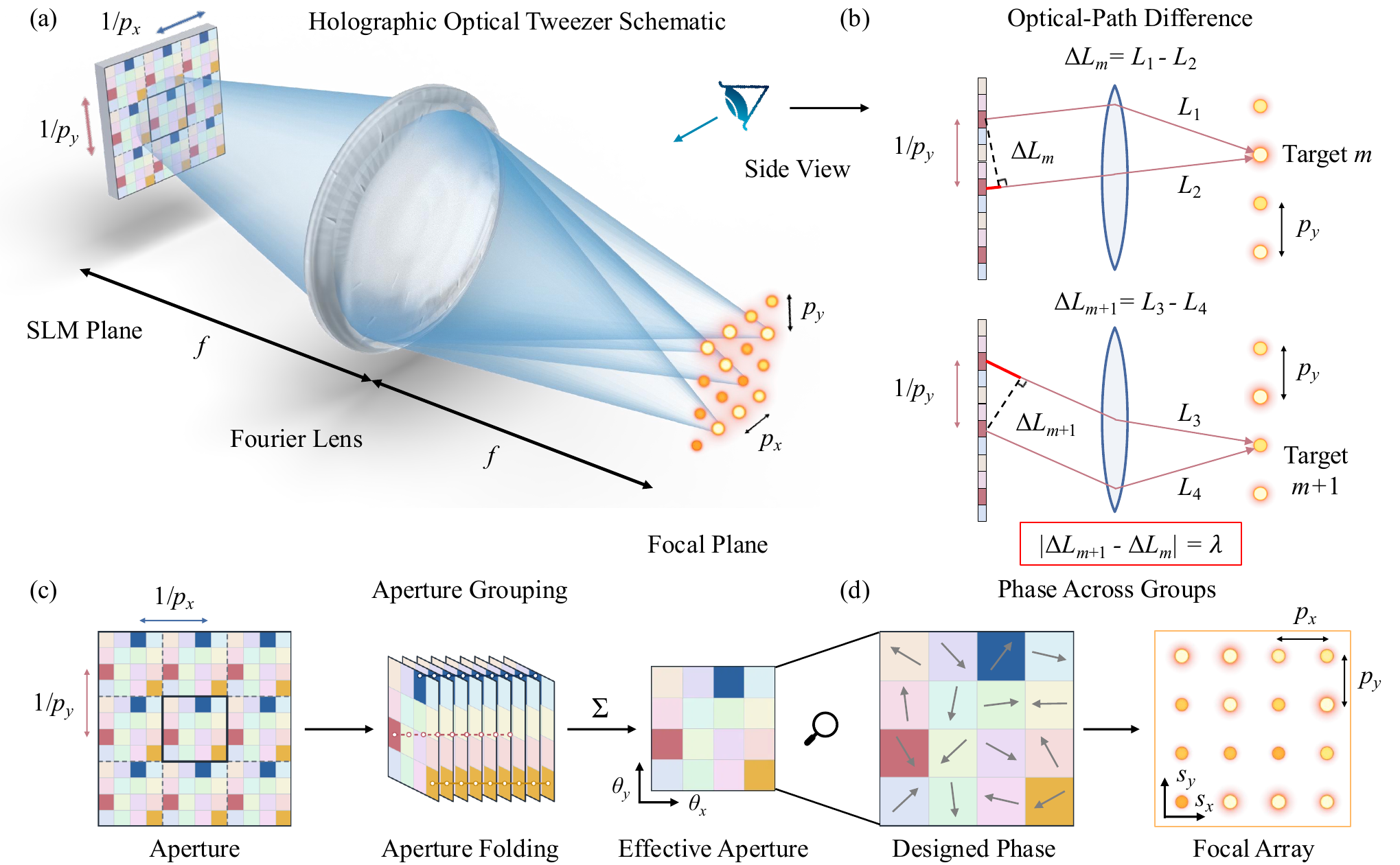}
\caption{Aperture grouping for a rectangular focal array. (a) Setup in the scaled transverse coordinates, with target spacings $p_x,p_y$. (b) Including the lens contribution, the optical-path difference for the illustrated positions changes by one wavelength between adjacent targets, preserving their relative phase. (c) Translating aperture cells onto one cell combines each group's complex fields into an effective aperture without changing the lattice fields. Colors identify groups set by the target lattice. The grouping need not align with SLM pixels. (d) Phase differences between groups distribute intensity among the targets.}
\label{fig:aperture-folding}
\end{figure*}

\subsection{How much power does the weakest trap need?}
\label{sec:optical-model}
We first relate the weakest-trap intensity to the required incident power. 
We use scalar paraxial propagation over a rectangular SLM aperture of widths $D_x,D_y$. Centered aperture coordinates $\mathbf x=(x,y)$ use these widths as units, so $\mathcal A=[-1/2,1/2]^2$. Focal coordinates $\mathbf s=(s_x,s_y)$ use units $\lambda f/D_x,\lambda f/D_y$, with wavelength $\lambda$ and effective focal length $f$. The beam amplitude $A(\mathbf x)$ is normalized to unit power and modulated by SLM phase $\phi(\mathbf x)$.

The focal field and intensity are \cite{goodman2017fourier}
\begin{equation}
\begin{aligned}
E(\mathbf s,\zeta)
&=\int_{\mathcal A}A(\mathbf x)e^{i\phi(\mathbf x)}
e^{-2\pi i\mathbf s\cdot\mathbf x}\\
&\quad\times e^{-i\pi\zeta[(D_x/D_y)x^2+(D_y/D_x)y^2]}\,\mathrm d^2x,\\
I(\mathbf s,\zeta)&=|E(\mathbf s,\zeta)|^2.
\end{aligned}
\label{eq:focal-field}
\end{equation}
For axial displacement $Z$ from the focal plane and incident power $P_{\rm in}$,
\begin{equation}
\zeta=\frac{D_xD_yZ}{\lambda f^2},\quad
I_{\rm phys}=\gamma P_{\rm in}I,\quad
\gamma=\frac{D_xD_y}{(\lambda f)^2}.
\label{eq:physical-intensity-scale}
\end{equation}
Here $I$ is dimensionless, $I_{\rm phys}$ is in $\mathrm{W\,m^{-2}}$, and $\gamma$ is the inverse focal-area scale. At $\zeta=0$, $E$ is the Fourier transform of the phase-modulated aperture field $A(\mathbf x)e^{i\phi(\mathbf x)}$. The quadratic phase accounts for defocus.

In the scalar dipole model, the attractive potential is proportional to negative local intensity \cite{grimm2000optical}. We therefore bound the intensity at each target and separately check whether it forms a confining local maximum through intensity curvature. The target geometry is specified by a lattice matrix $\mathbf B$, offset $\mathbf c$ and $K=|\mathcal T|$ integer labels in $\mathcal T$. The matrix columns are two independent lattice vectors. The target positions, weakest-target objective and incident power required for physical intensity $I_{\rm req}$ at every target obey
\begin{equation}
\begin{aligned}
\mathbf s_{\mathbf n}&=\mathbf c+\mathbf B\mathbf n,
\qquad \mathbf n\in\mathcal T,\\
I_{\min}(\phi)&=\min_{\mathbf n\in\mathcal T}I(\mathbf s_{\mathbf n},0),
\qquad J(\phi)=K I_{\min}(\phi),\\
P_{\rm in}&=\frac{I_{\rm req}}{\gamma I_{\min}(\phi)}
=\frac{K I_{\rm req}}{\gamma J(\phi)}.
\end{aligned}
\label{eq:array-objective}
\end{equation}
Thus maximizing $J$ minimizes the input power needed to meet the intensity threshold at every site.
For equal target intensities, $J$ equals their point-intensity sum. Rectangular spacings give $\mathbf B=\operatorname{diag}(p_x,p_y)$, and the same notation covers rotated, sheared and triangular arrays.

\subsection{How do illumination and geometry limit every trap?}
\label{sec:finite-device-bounds}
\paragraph{Aperture grouping gives a phase-independent bound.}
The main observation is that some aperture positions contribute with the same relative phase to every target in the lattice. We combine these positions into groups, reducing the array problem to an effective aperture.
For a rectangular array, separations $1/p_x$ along $x$ or $1/p_y$ along $y$ give one additional full turn of relative phase between neighboring targets (\cref{fig:aperture-folding}). More generally, an aperture displacement $\mathbf B^{-T}\mathbf r$, with integer $\mathbf r$, contributes $e^{-2\pi i\mathbf n\cdot\mathbf r}=1$. Aligning a group at one target therefore aligns it throughout the lattice. The phase assigned to each group controls interference between groups. The illumination sets the amplitudes being added, while lattice spacing and orientation determine which positions are combined.

To quantify this limit, we fold the aperture cells onto one reference cell with group coordinate $\boldsymbol\theta\in\mathcal Q=[-1/2,1/2)^2$. This combines each group's contributions without changing the lattice fields, which become Fourier coefficients of the effective aperture. After compensating the offset phase, aligned contributions attain the summed amplitude $\overline A_{\mathbf B}$. The triangle inequality and Parseval then bound the lattice point-intensity sum, and hence the target minimum:
\begin{equation}
\begin{aligned}
\overline A_{\mathbf B}(\boldsymbol\theta)
&=\frac1{|\det\mathbf B|}\sum_{\mathbf r\in\mathbb Z^2}
A\!\left(\mathbf B^{-T}(\mathbf r+\boldsymbol\theta)\right),\\
J(\phi)&\le C_A(\mathbf B)
=\int_{\mathcal Q}\overline A_{\mathbf B}^{\,2}\,\mathrm d^2\theta.
\end{aligned}
\label{eq:lattice-intensity-bound}
\end{equation}
Terms outside the aperture vanish, and $1/|\det\mathbf B|$ accounts for the cell-area scaling. 
Note that the bound $C_A$ depends on illumination and lattice geometry, but not on the chosen phase pattern. 
Aligning each group avoids cancellation within that group; every global optimum with unrestricted spatial phase does so for almost every illuminated group (Supplement~1, Sec.~SII). 
Reaching the bound also requires the groups to direct their combined intensity to the targets and distribute it evenly. Thus coherent alignment alone is necessary but not sufficient.

\paragraph{SLM pixels impose an additional limit.}
On an $N_x\times N_y$ SLM, the phase is uniform within each pixel, and each pixel's phase is independently adjustable over a continuous range.
For unit fill factor and constant incident amplitude within each rectangular pixel, target $k$ has intensity \cite{arrizon1997efficiency,arrizon1999implementation}
\begin{equation}
\begin{aligned}
I(\mathbf s_k,0)&=d_k|F_k|^2,\qquad
d_k=\prod_{d=x,y}\operatorname{sinc}^2(s_{k,d}/N_d),\\
F_k&=\frac1P\sum_j A_j e^{i\phi_j-2\pi i\mathbf s_k\cdot\mathbf x_j}.
\end{aligned}
\label{eq:target-pixel-response}
\end{equation}
Here $P=N_xN_y$ and pixel $j$ has center $\mathbf x_j$, amplitude $A_j$ and phase $\phi_j$. We use $\operatorname{sinc}u=\sin(\pi u)/(\pi u)$, continued at zero. Pixel phases control interference between pixels, while $d_k$ describes cancellation within each pixel that a constant phase cannot correct. Smaller $d_k$ therefore requires larger $|F_k|^2$ for the same target intensity. Shifting the array changes this cost through the absolute target positions in $d_k$. With continuous spatial phase control, a linear ramp shifts the array without changing $C_A(\mathbf B)$.

Let $J^\star$ be the optimum over the pixel phases. If any $d_k=0$, its target remains dark and $J^\star=0$. For $d_k>0$ at every target, the same pixel phases must serve all traps, coupling the individual pixel-response requirements. For $\mathbf N=(N_x,N_y)$ and positive weights $\omega_j$, a weighted overlap matrix captures correlations between target propagation phases across the SLM and bounds their combined Fourier-sum intensity:
\begin{gather}
\beta_{\mathbf N}=\frac1P\sum_j\frac{A_j^2}{\omega_j},\quad
(\mathbf G_{\mathbf N})_{kl}
=\frac1P\sum_j\omega_j e^{-2\pi i(\mathbf s_k-\mathbf s_l)\cdot\mathbf x_j},
\label{eq:weighted-gram-matrix}\\
I_{\min}\sum_{k=1}^K d_k^{-1}
\le\sum_{k=1}^K|F_k|^2
\le\beta_{\mathbf N}\lambda_{\max}(\mathbf G_{\mathbf N}).
\label{eq:weighted-sampling-bound}
\end{gather}
Combining \cref{eq:lattice-intensity-bound,eq:weighted-sampling-bound} gives
\begin{equation}
J(\phi)\le J^\star\le U_{\mathbf N}
=\min\left\{C_A(\mathbf B),
\frac{\beta_{\mathbf N}\lambda_{\max}(\mathbf G_{\mathbf N})}
{K^{-1}\textstyle\sum_{k=1}^K d_k^{-1}}\right\}.
\label{eq:finite-device-upper}
\end{equation}
The mean inverse pixel response measures the cost of compensating for diffraction across the array. 
The bound includes the classical array-illuminator result \cite{arrizon1997efficiency,arrizon1999implementation} as a special case (Supplement~1, Sec.~SII).
Importantly, for a hologram with measured or calculated $J(\phi)$, further phase optimization can save no more than $1-J(\phi)/U_{\mathbf N}$ of its incident power.
This turns an intensity bound into a practical test of whether better phase optimization can still help.

\subsection{Can phase-only holograms reach the intensity limit?}
An upper bound is most useful if we know whether it can be reached. We therefore determine the attainable limit as target and pixel counts grow together.
The targets occupy all lattice sites in an expanding region of fixed shape. We denote its linear scale by $M$ and its number of sites by $K_M$.  The aperture, illumination, lattice and offset remain fixed. Pixel counts grow in proportion to array width, so targets occupy a fixed fraction of the sampling range and pixel diffraction remains relevant. We assume that every aperture group receives light, that the target band stays separated from its sampling replicas as the array grows, and that the single-pixel response stays nonzero across the band. Under the full conditions of Supplement~1, Sec.~SIII, which cover uniform and Gaussian illumination, we obtain
\begin{equation}
\begin{aligned}
\lim_{M\to\infty}J_M^\star&=\frac{C_A(\mathbf B)}{\mathcal H_\infty},\\
\mathcal H_\infty&=\lim_{M\to\infty}\frac1{K_M}
\sum_{k=1}^{K_M}d_{k,\mathbf N(M)}^{-1}.
\end{aligned}
\label{eq:general-planar-limit}
\end{equation}
Here $\mathcal H_\infty$ is the limiting mean inverse pixel response from \cref{eq:target-pixel-response}. This matching law shows that the weakest-trap intensity scales as $1/K_M$, with a coefficient set by aperture grouping and the average inverse pixel response. The upper limit is therefore asymptotically achievable under the stated assumptions, not merely a loose estimate. Continuous spatial phase gives $\mathcal H_\infty=1$.

To approach the limit, the construction first aligns the aperture groups. It then allocates more pre-diffraction intensity to targets with weaker pixel responses, so their final intensities become equal. Coordinate-mapping and optimal-transport beam shaping \cite{romero1996lossless,aagedal1997analytical,swan2025high,torchylo2026fast} implement this allocation: the effective-aperture phase gradient selects target directions, while the phase curvature controls the aperture area contributing coherently to each target. Coordinated pixel phases reproduce these fields under the original illumination, with the added field vanishing uniformly relative to the target fields. With illumination-dependent weights, the upper bound converges to the same intensity.

High intensity at the requested coordinates does not by itself guarantee stable traps. We therefore also examine local focusing in three dimensions. 
A transverse displacement adds a linear aperture phase, while defocus adds a quadratic phase. Their variation within an aligned group determines local intensity curvature. If the limiting intensity relative to its target value decreases quadratically away from each target in every direction with a uniform curvature margin, the masks attain the same intensity limit with distinct strict three-dimensional maxima near all targets, vanishing displacements and target-to-peak intensity ratios tending to one (Supplement~1, Sec.~SIV).

\begin{figure}[!t]
\centering
\includegraphics[width=\columnwidth]{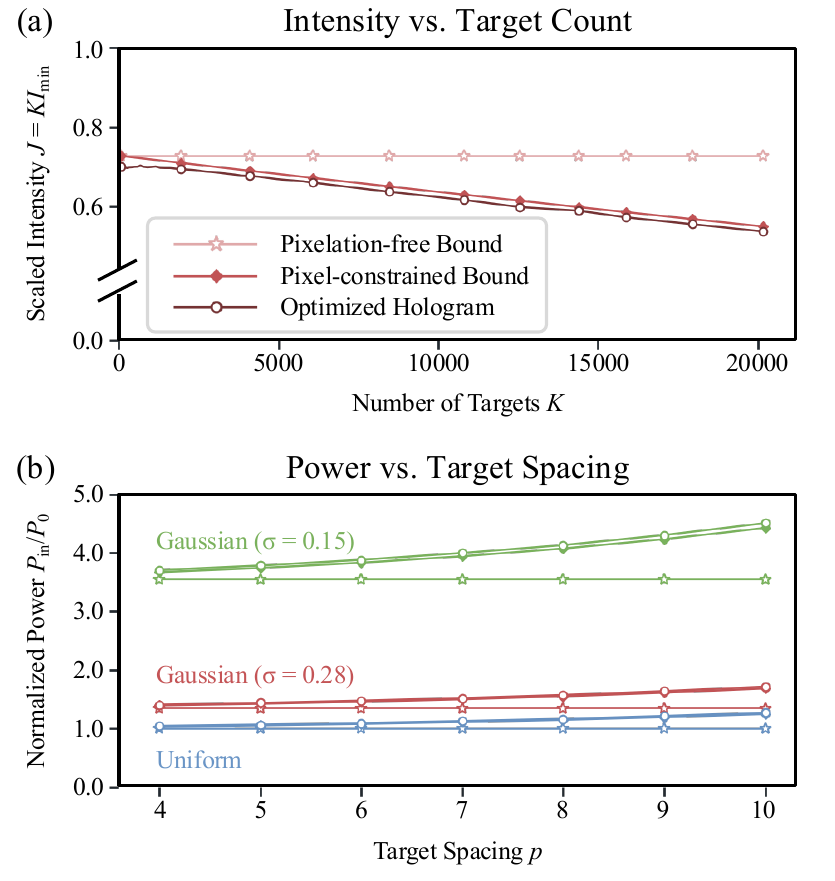}
\caption{Planar arrays generated using $1024\times1024$ SLM pixels. (a) $J=K I_{\min}$ and bounds $C_A(\mathbf B),U_{\mathbf N}$ for 43 rectangular arrays spanning 63 to 20,163 targets, with noninteger spacing, nonzero offset and Gaussian-shaped illumination ($\sigma=0.28$). The vertical axis omits $0.2<J<0.5$. (b) Required input power and the corresponding pixel-constrained and pixelation-free lower bounds for $63\times63$ targets, with the same markers as in (a). Power, spacing and $\sigma$ use units $P_0=K I_{\mathrm{req}}/\gamma$, $\lambda f/D$ and $D$, respectively, for square-aperture width $D$. Profiles are specified in Supplement~1, Sec.~SVI.}
\label{fig:planar-validation}
\end{figure}

\subsection{How close are practical holograms to the limit?}
On a fixed SLM, numerically optimized holograms reach $95.93\%$--$98.47\%$ of their pixel-constrained bounds (Fig.~\ref{fig:planar-validation}(a)). As the array expands, single-pixel diffraction weakens the response at its outer targets, reducing $U_{\mathbf N}$ while $C_A(\mathbf B)$ stays constant. Across the sweep from 63 to 20,163 targets, $J$ falls by $23.3\%$ and remains close to $U_{\mathbf N}$. At 20,163 targets, the bounds limit additional input-power savings to $2.27\%$ with the existing pixels and $26.12\%$ with continuous spatial phase control of the same illumination. Supplement~1, Secs.~SVI~A--B, gives the endpoint values, numerical methods and local-focus checks.

The bounds also reveal which design choices can reduce power when phase optimization is nearly exhausted. 
Figure~\ref{fig:planar-validation}(b) compares input-power requirements across illuminations and target spacings. Illumination sets the amplitudes added within each group, while increasing spacing moves outer traps farther into the pixel-diffraction envelope, raising the required power over this range. \mbox{For $\sigma=0.28$}, a $p=4$ hologram meets the intensity requirement at about $1.41P_0$, whereas $p=10$ requires at least about $1.68P_0$ for any hologram.

\subsection{Can splitting the array across two planes save power?}

A way to reduce pixel-diffraction loss is to distribute a fixed number of traps across depth, reducing the transverse extent of each plane. But this introduces a new challenge: one hologram must balance different defocus responses, and a large total intensity can still leave one plane weak. We consider two $M\times M$ arrays at depths $0$ and $\delta$, under uniform square illumination, both with lateral spacing $p\in\mathbb N$ and center $\mathbf s=0$. The group responses $b_0,b_\delta$ average the same $p^2$ aperture phase factors $v_{\mathbf r}$, without and with the defocus phase in \cref{eq:focal-field}. Define
\begin{equation}
C_2(p,\delta)=\tfrac12\int_{\mathcal Q}\max\nolimits_{|v_{\mathbf r}|=1}
\bigl(|b_0|^2+|b_\delta|^2\bigr)\,\mathrm d^2\theta.
\label{eq:two-plane-group-capacity}
\end{equation}
Thus $2C_2$ bounds the point-intensity sum over both complete lattices. Unlike the planar bound, it explicitly accounts for the competition between the two axial responses.

To show how the two planes can share intensity fairly, phase reversal and defocus compensation produce a second pattern with responses $(\overline b_\delta,\overline b_0)$, exchanging the field magnitudes between planes. In each group, a unit-modulus factor $z_{\rm q}$ puts the paired fields in quadrature, cancelling intensity cross terms in both planes:
\begin{equation}
|b_0+z_{\rm q}\overline b_\delta|^2
=|b_\delta+z_{\rm q}\overline b_0|^2
=|b_0|^2+|b_\delta|^2.
\label{eq:two-plane-balance}
\end{equation}
One phase-only hologram carries both patterns on the SLM pixel grid in alternating narrow bands. The phase across the bands ensures coherent addition at every target in both planes.

\begin{figure}[!t]
\centering
\includegraphics[width=\columnwidth]{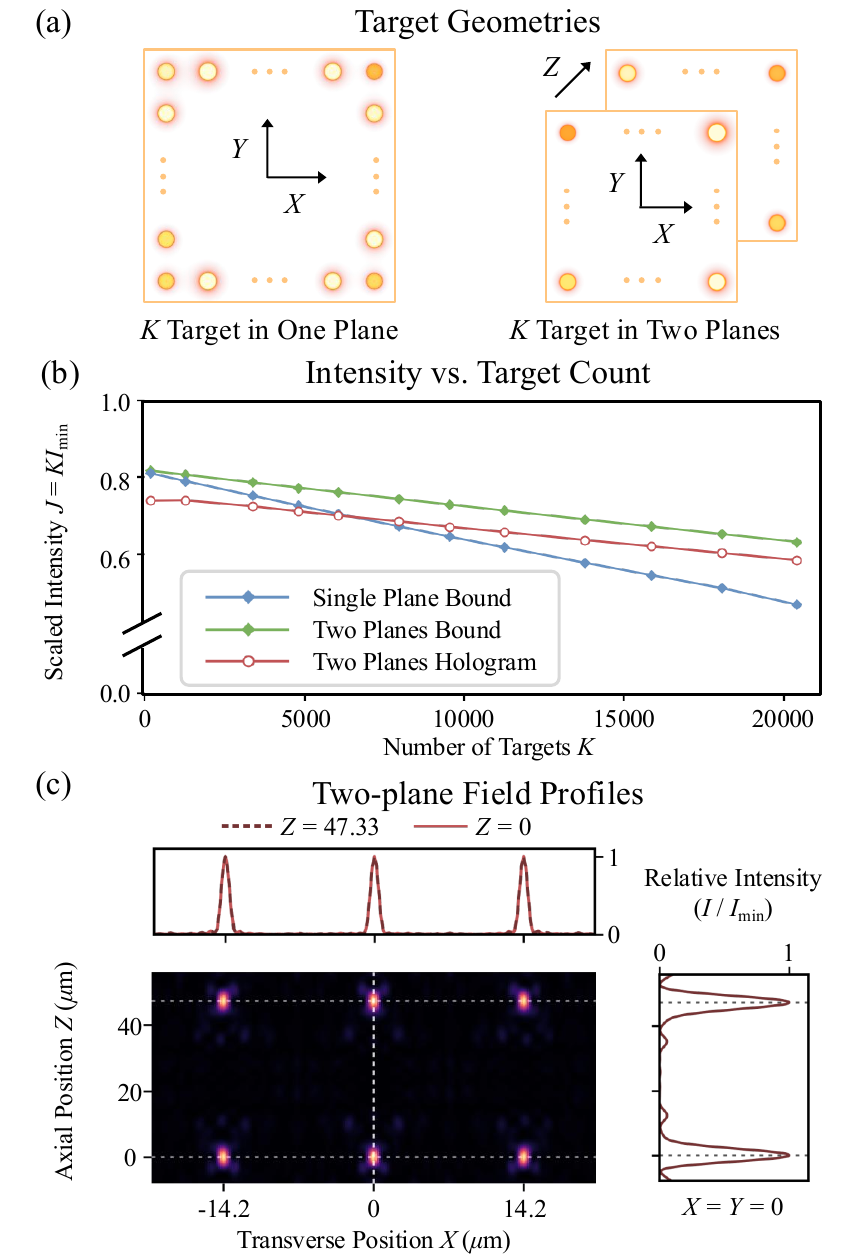}
\caption{Numerical comparison on a modeled $2160\times2160$-pixel SLM with uniform $x$-polarized illumination at $852\,\mathrm{nm}$. Lateral spacing is $14.20\,\mu\mathrm m$ and plane separation is $47.33\,\mu\mathrm m$. (a) Rectangular one-plane and two-plane layouts at equal total target count and lateral spacing. (b) Minimum target intensity times total trap count for optimized two-plane holograms, with the planar and two-plane upper bounds. The vertical scale is compressed below $J=0.45$. (c) Central $X$--$Z$ intensity section at $Y=0$ for the $7,938$-target hologram, with transverse profiles at both target depths above and an axial profile at $X=Y=0$ alongside. The intensity reference in (c) is the minimum at the targets across both planes.}
\label{fig:two-plane-arrays}
\end{figure}

Here $I_{\min}$ is the minimum over both planes, with $K=K_M=2M^2$. Axial separation limits coherent overlap, while sampling resolves defocus. For odd $M$ on an $N\times N$ pixel grid, one sufficient regime is $\delta\ge64$, $p\ge2\delta$, and $N/M\to\nu>p$, giving the scalar matching law (Supplement~1, Sec.~SV)
\begin{equation}
\lim_{M\to\infty}J_M^\star
=\frac{2C_2(p,\delta)}{\mathcal H_\infty}.
\label{eq:two-plane-limit}
\end{equation}
The two planes share the same limiting pixel response because their transverse coordinates coincide. In this scalar regime, the attaining masks also form a distinct strict three-dimensional intensity maximum near each target for sufficiently large arrays.

The depth exchange carries over to vector Debye--Wolf propagation \cite{richards1959electromagnetic}, where the polarization transport factors are real and one pixel phase multiplies all field components. Let $2C_{2,A}^{\mathrm{vec}}$ be the integral over the group cell of the maximum total vector intensity at both depths using a common set of phases.

\begin{samepage}
The construction needs sufficiently independent responses at the two depths within each aperture group to balance their intensities. For the square aperture modeled in Fig.~\ref{fig:two-plane-arrays} (direction-cosine half-width $a=0.45$), this independence holds uniformly for aligned arrays at pitch $p=15$ throughout $36\le\delta\le54$, giving a sufficient regime for matching (Supplement~1, Sec.~SV~I).

As odd $M$ grows with $N/M\to\nu>p$, the optimum at the specified target positions obeys
\begin{equation}
\lim_{M\to\infty}J_M^\star
=\frac{2F_{\rm pix}^2 C_{2,A}^{\mathrm{vec}}}
       {\mathcal H_\infty^{\rm act}} .
\label{eq:two-plane-vector-limit}
\end{equation}
\par
\end{samepage}
The active area fraction $F_{\rm pix}$ scales the on-axis field of each pixel, giving $F_{\rm pix}^2$ in intensity. The factor $\mathcal H_\infty^{\rm act}$ averages the inverse intensity response relative to on axis across the limiting target band. This result uses centered square active regions, smooth positive illumination, fixed linear $x$ polarization, and continuously adjustable pixel phases.

On a fixed SLM, Fig.~\ref{fig:two-plane-arrays} compares optimized two-plane holograms with 8-bit phase quantization to the planar and two-plane intensity bounds. The vector calculation uses the same illumination and pixel grid for both layouts, with $p=15$ and a two-plane separation of $\delta=45$. The planar bound covers every equal-count selection and common translation on the same pixel-aligned, fixed-pitch lattice at nominal focus within the central sampling window.

The lower intensity estimates for the twelve holograms are $90.5\%$--$92.5\%$ of the corresponding two-plane upper bounds. Numerical focus checks find a strict three-dimensional intensity maximum near every target, with intensity at the target at least $99\%$ of its local peak. For the same minimum target intensity, the comparison certifies a power saving over the planar family at the seven largest sampled counts, beginning at 7,938 targets and reaching at least $19.68\%$ at 20,402 targets (Supplement~1, Sec.~SVI~E).

\subsection{What the bounds mean for optical-tweezer design}
The central practical result is a way to separate two sources of power savings: better phase patterns and changes to the optical system. Once a hologram approaches its pixel-constrained bound, phase optimization alone offers little additional benefit.
Figure~\ref{fig:planar-validation}(b) shows how illumination alters coherent group amplitudes and spacing changes grouping and pixel responses, hence the required power. Splitting a fixed target count between planes reduces transverse extent and deflection-dependent diffraction loss \cite{manetsch2025tweezer}, while one hologram must balance coherent depth responses (Fig.~\ref{fig:two-plane-arrays}). At the larger sampled counts, the two-plane holograms exceed the bound for every equal-count planar layout in the stated fixed-pitch family. Their gain survives depth coupling and cannot be recovered by phase optimization within that family.
The demonstrated two-plane power advantage can complement the packing and transport benefits of three-dimensional qubit layouts \cite{wu2026efficient}. Extending the comparison to larger stacks would determine how interference between planes changes the power advantage from reduced transverse extent.

Beyond target geometry, polarization provides additional control over the focal field. Our vector calculation includes the polarization-dependent intensity distribution and longitudinal fields produced by tight focusing \cite{dorn2003sharper}, while keeping the incident linear polarization fixed. Spatially varying polarization \cite{bhebhe2018vector,sun2026local} allows a broader range of focal fields. 
Extending the bounds to spatially varying polarization would quantify whether this additional control can lower the required incident power. More broadly, the bounds provide a benchmark for deciding when to improve the hologram and when to change illumination, target layout, sampling, or available optical degrees of freedom.

\paragraph*{Acknowledgment} K.~Y.~W. received partial support through the University of Washington's Accelerating Quantum-Enabled Technologies (AQET) traineeship under National Science Foundation Award No.~DGE-2021540.

\paragraph*{Disclosures} The authors declare no conflicts of interest.

\paragraph*{Data availability}
The code will be made publicly available.

\paragraph*{Supplemental document} Supplement~1 has supporting content.

\bibliographystyle{references}
\bibliography{references}
\end{document}